\documentclass[12pt]{article}
\usepackage{amsmath, amssymb, slashed, epsf, color}
\usepackage{graphicx}

\def\lsim{\mathrel{\raise.3ex\hbox{$<$\kern-.75em\lower1ex\hbox{$\sim$}}}}
\def\gsim{\mathrel{\raise.3ex\hbox{$>$\kern-.75em\lower1ex\hbox{$\sim$}}}}

\begin{document}

\begin{titlepage}
\begin{center}

\hfill IPMU-12-0232 \\
\hfill \today

\vspace{1.5cm}
{\large\bf Model Independent Analysis of Interactions \\
between Dark Matter and Various Quarks}
\vspace{2.0cm}

{\bf Biplob Bhattacherjee}$^{(a)}$,
{\bf Debajyoti Choudhury}$^{(b)}$,
{\bf Keisuke Harigaya}$^{(a)}$,
\\
{\bf Shigeki Matsumoto}$^{(a)}$
and
{\bf Mihoko M. Nojiri}$^{(c, a)}$

\vspace{1.0cm}
{\it
$^{(a)}${Kavli IPMU, University of Tokyo, Kashiwa, 277-8583, Japan} \\
$^{(b)}${Department of Physics and Astrophysics, \\
University of Delhi, Delhi 110007, India.} \\
$^{(c)}${\it Theory Group, KEK, Tsukuba, 305-0801, Japan}
}
\vspace{2.0cm}

\abstract{ Present and future expected limits on interactions between
  dark matter and various quarks are thoroughly investigated in a
  model-independent way. In particular, the constraints on the
  interactions from the Large Hadron Collider (LHC) experiment are
  carefully considered with a focus on mono jet + missing transverse
  energy ($\slashed{E}_T$), mono b-jet + $\slashed{E}_T$, and top
  quark(s) + $\slashed{E}_T$ channels. Model-independent upper limits
  (expected limits) on the cross section times acceptance for
  non-standard model events are derived for the LHC operating 
  at 7/8/14 TeVs. Assuming that the dark matter is a
  singlet real scalar or a singlet Majorana fermion, we also put
  constraints on several operators describing its interactions with
  up, down, strange, charm, bottom and top quarks. These constraints
  are compared to those obtained from cosmological and astrophysical
  implications.}

\end{center}
\end{titlepage}
\setcounter{footnote}{0}

\section{Introduction}
\label{sec: introduction}

The origin of the dark matter in the Universe and its nature
constitute some of the most outstanding mysteries in modern particle
physics, cosmology, and astrophysics. Though the problem is not
resolved yet, many efforts have been made over the years (and continue
to be made) in the search for dark matter (DM), and these give
precious information about several DM interactions. Theoretically, the
weakly interacting massive particle (WIMP), called so as its couplings
are of the order of the electroweak (EW) couplings, is one of the
promising candidates for dark matter~\cite{Bertone:2010zz}.
Experimentally, direct detection experiments of DM put very severe
limits on the scattering cross section between the DM and a
nucleon~\cite{:2012nq}, while indirect detection experiments give
constraints on self-annihilation cross sections of the DM into various
final states~\cite{Indirect}.

Among DM interactions, those to quarks and gluons are particularly
well investigated thanks to the Large Hadron Collider (LHC)
experiment. Importantly, the LHC experiment enables us to explore not
only the interactions relevant to direct and indirect DM detections
but also certain interactions irrelevant to those
detections. Consequently, many studies have already been performed
assuming a definite relation among the DM interactions with various
quarks or using an analysis with simple collider
simulations~\cite{Past works}. In this article, we thoroughly
investigate present and (near) future expected limits on the DM
interaction with individual quark flavours%
\footnote{The DM interaction with gluon can also be investigated by the
LHC experiment and severe constraints are obtained in past works~\cite{Past works}.}, focussing on mono jet +
missing transverse energy ($\slashed{E}_T$), mono b-jet +
$\slashed{E}_T$, and top quark(s) + $\slashed{E}_T$ channels, which
are all derived from dedicated collider simulations using a
model-independent method.

In the next section (section \ref{sec: operators}), we first summarize
DM interactions assuming that the DM particle is a real scalar or a
Majorana fermion and singlet under SM gauge groups\footnote{The
  requirement of being singlet is not strictly necessary, but a
  simplifying assumption.  A charged DM in the sky would interact with
  cosmic rays (impeding their propagation), as well as the CMB photons
  (thereby distorting the blackbody spectrum beyond permissible
  limits). Similarly, neutral but coloured DM would be captured by
  nuclei to lead to exotic isotopes. A prominent counterexample of a
  neutral, but gauge nonsinglet DM is given by the lightest neutralino
  in the minimal supersymmetric standard model.}. Such candidates are
frequently used in the literature and can be regarded as the simplest
examples of a DM particle. We then put limits on the operators from
cosmological and astrophysical observations of the DM in section
\ref{sec: cosmo_astro}. We next consider what kinds of DM signals are
generally expected at the LHC experiment in section \ref{sec: LHC} and
give model-independent upper limits (expected limits) on the cross
section times acceptance for DM signals at 7 TeV (8 or 14 TeV) run. It
turns out that these limits are very useful to constrain any DM quark
interactions. Using the obtained limits, we also explain how to put a
limit on each operator (discussed in section \ref{sec: operators})
describing a DM interaction with various quarks. We finally consider,
in
section \ref{sec: results}, 
how severe limits can the LHC experiment  put 
onsuch operators  and compare  with those obtained
from cosmological and astrophysical observations. Section
\ref{sec: summary} is devoted to a summary of our discussions.

\section{Dark matter interactions}
\label{sec: operators}

Dark matter interactions with various quarks are discussed in this
section assuming that the DM is a real scalar or a Majorana fermion
which is singlet under SM gauge groups. We introduce operators
describing interactions between the DM and SM particles up to 
mass-dimension six. Decoupling and/or weakly coupled heavy physics
behind the interactions are implicitly assumed, so that the whole
theory (i.e., the ultraviolet completion) is renormalizable. As the
stability of the DM is guaranteed by imposing a $Z_2$ symmetry (under
which the DM field is odd, while all the SM fields are even), the
operators have to involve two DM fields. The real scalar DM is denoted
by $\phi$, while the Majorana fermion DM is denoted by $\chi$ in the
following discussions.

The effective lagrangian for the scalar DM field $\phi$ is simply given by
\begin{eqnarray}
{\cal L}_{\rm eff}^{(\phi)}
=
{\cal L}_{\rm SM}
+ \left[
\frac{1}{2} \left( \partial \phi \right)^2
- \frac{M_\phi^2}{2} \phi^2
\right]
+ \sum_{n = 4}^\infty \frac{1}{\Lambda^{n - 4}}
\left[
\sum_i c_i \, {\cal O}_i^{(n)}
+ h.c.
\right],
\end{eqnarray}
where ${\cal L}_{\rm SM}$ is the SM lagrangian. Since several
interactions contribute to the DM mass after the electroweak symmetry
breaking, the physical DM $m_\phi$ (we use this notation throughout
the paper) is distinct from $M_\phi$. The mass dimension of operator
${\cal O}_i^{(n)}$ is denoted by $n$, while $\Lambda$ is the cutoff
scale below which the effective lagrangian describes the physics. 
The complete set of operators upto dimension six, which are
invariant under both Lorentz group and SM gauge groups, is given by
\begin{eqnarray}
\begin{array}{ll}
{\cal O}^{(4)}_{\phi H} = \phi^2 |H|^2, &
{\cal O}^{(6)}_{\phi L H E} = \phi^2 \bar{L} H E, \\
{\cal O}^{(6)}_{\phi H} = \phi^2 |H|^4, &
{\cal O}^{(6)}_{\phi Q H D} = \phi^2 \bar{Q} H D, \\
{\cal O}^{(6)}_{\phi \partial H} = \phi^2 |D_\mu H|^2, &
{\cal O}^{(6)}_{\phi Q H U} = \phi^2 \bar{Q} H^c U, \\
{\cal O}^{(6)}_{\phi VV} = \phi^2 V_{\mu\nu} V^{\mu\nu}, &
{\cal O}^{(6)}_{\phi V \tilde{V}} = \phi^2 V_{\mu\nu} \tilde{V}^{\mu\nu}, \\
\end{array}
\end{eqnarray}
where $H$, $L = (\nu, e_L)^T$, $E$, $Q = (u_L, d_L)^T$, $D$, and $U$
are the higgs doublet, the lepton doublet, the charged lepton singlet,
the doublet quark, the down-type singlet quark, and the up-type
singlet quark, respectively. On the other hand, $V_{\mu\nu}$
($\tilde{V}_{\mu\nu} \equiv \epsilon_{\mu\nu\alpha\beta} V^{\alpha
\beta}$) is the field strength tensor for a gauge boson, namely,
$V_{\mu\nu} = B_{\mu\nu}$, $W_{\mu\nu}^j$, and $G_{\mu\nu}^a$ are
those for the hyper-charge gauge boson, the weak SU(2)$_L$ gauge boson,
and the gluon, respectively. The covariant derivative is denoted by
$D_\mu$. Operators ${\cal O}^{(6)}_{\phi L H E}$, ${\cal
O}^{(6)}_{\phi Q H D}$, and ${\cal O}^{(6)}_{\phi Q H U}$ are in fact
flavor-dependent (e.g., $\phi^2 \bar{Q}_I H D_J)$. In order to avoid
dangerous flavor changing processes, we only consider flavor diagonal
parts of the operators\footnote{This is not strictly necessary,
though. For example, the leading flavour changing neutral current
interaction that a term such as $\phi^2 \, \bar Q_3 \, H^c \, u_1$
would generate is the anomalous $\bar t \, u \, H$ vertex, and that
too at one-loop. This clearly is harmless. Nonetheless, we desist from
admitting such terms, simply because their inclusion does not result
in any qualitative change in DM physics.}. In other words, we 
implicitly assume that heavy physics has some mechanism to suppress
flavor off-diagonal parts of the operators.

Since we are interested in interactions between DM and various quarks,
we focus on ${\cal O}^{(6)}_{\phi Q_i H D_i}$ and ${\cal
  O}^{(6)}_{\phi Q_i H U_i}$ ($i = 1, 2, 3$) among the operators
listed above. After the electroweak symmetry breaking ($\langle 0
|H|0\rangle = (0, v)^T/\sqrt{2}$ with $v$ being about 246 GeV), these
operators give following effective interactions,
\begin{eqnarray}
{\cal L}_{\rm int}^{(\phi)}
=
\frac{v\phi^2}{\sqrt{2} \Lambda^2} \sum_{i = 1}^3
\left[
\bar{u}_i
\left(c_{\phi Q_i H U_i}^{(R)} + i c_{\phi Q_i H U_i}^{(I)} \gamma_5 \right)
u_i
+
\bar{d}_i
\left(c_{\phi Q_i H D_i}^{(R)} + i c_{\phi Q_i H D_i}^{(I)} \gamma_5 \right)
d_i
\right],
\label{eq: scalar int}
\end{eqnarray}
where $c_{\phi Q_i H U_i}^{(R)}$ ($c_{\phi Q_i H U_i}^{(I)}$) and
$c_{\phi Q_i H D_i}^{(R)}$ ($c_{\phi Q_i H D_i}^{(I)}$) are the real
(imaginary) parts of the Wilson coefficients $c_{\phi Q_i H U_i}$ and
$c_{\phi Q_i H D_i}$ corresponding to the operators ${\cal
O}^{(6)}_{\phi Q_i H U_i}$ and ${\cal O}^{(6)}_{\phi Q_i H D_i}$.

As in the case of the real scalar DM field $\phi$, the effective
lagrangian for the Majorana fermion DM field $\chi$ is also given in
the same form,
\begin{eqnarray}
{\cal L}_{\rm eff}^{(\chi)}
=
{\cal L}_{\rm SM}
+ \frac{1}{2}
\bar{\chi} \left(i\slashed{\partial} - M_\chi \right) \chi
+ \sum_{n = 5}^\infty \frac{1}{\Lambda^{n - 4}}
\left[
\sum_i c_i \, {\cal O}_i^{(n)}
+ h.c.
\right],
\end{eqnarray}
where we adopt the four component notation for the Majorana field $\chi
= \chi^c$ with the superscript `$c$' denoting charge
conjugation. Once again, the physical mass $m_\chi$ is distinct 
from the mass parameter 
$M_\chi$. The complete set of effective operators involving two
DM fields is given by
\begin{eqnarray}
\begin{array}{ll}
{\cal O}^{(5)}_{\chi H, 1} = (\bar{\chi} \chi) |H|^2, &
{\cal O}^{(6)}_{\chi U} = (\bar{\chi} \gamma^\mu \gamma_5 \chi)
(\bar{U} \gamma_\mu U) \\
{\cal O}^{(5)}_{\chi H, 2} = i(\bar{\chi} \gamma_5 \chi) |H|^2, &
{\cal O}^{(6)}_{\chi D} = (\bar{\chi} \gamma^\mu \gamma_5 \chi)
(\bar{D} \gamma_\mu D) \\
{\cal O}^{(6)}_{\chi H} = (\bar{\chi} \gamma^\mu \gamma_5 \chi)
(H^\dagger i\overleftrightarrow{D_\mu} H), &
{\cal O}^{(6)}_{\chi L} = (\bar{\chi} \gamma^\mu \gamma_5 \chi)
(\bar{L} \gamma_\mu L) \\
{\cal O}^{(6)}_{\chi Q} = (\bar{\chi} \gamma^\mu \gamma_5 \chi)
(\bar{Q} \gamma_\mu Q), &
{\cal O}^{(6)}_{\chi E} = (\bar{\chi} \gamma^\mu \gamma_5 \chi)
(\bar{E} \gamma_\mu E). \\
\end{array}
\end{eqnarray}
For the same reason as in the scalar DM case, we focus only on flavor
diagonal parts of the operators ${\cal O}^{(6)}_{\chi Q}$, ${\cal
  O}^{(6)}_{\chi U}$, and ${\cal O}^{(6)}_{\chi D}$, which lead to 
\begin{eqnarray}
{\cal L}_{\rm int}^{(\chi)}
=
\frac{2\bar{\chi} \gamma^\mu \gamma_5 \chi}{\Lambda^2} \sum_{i = 1}^3
\left[
\bar{u}_i \gamma_\mu \left(c_{\chi U_i}P_R + c_{\chi Q_i}P_L \right)u_i
+
\bar{d}_i \gamma_\mu \left(c_{\chi D_i}P_R + c_{\chi Q_i}P_L \right)d_i
\right],
\label{eq: fermion int}
\end{eqnarray}
where all coefficients $c_{\chi U_i}$, $c_{\chi D_i}$, and $c_{\chi
Q_i}$ are real numbers.

\section{Cosmological and astrophysical constraints}
\label{sec: cosmo_astro}

We are now ready to put limits on the Wilson coefficients $c_i$.  In
this section, we consider various limits obtained from cosmological
and astrophysical observations, while those obtainable from the LHC
experiments will be discussed in the next section. For the real scalar
DM $\phi$, twelve operators, {\rm viz.}  $\phi^2 \bar{u}_i u_i$,
$\phi^2 \bar{u}_i i\gamma_5 u_i$, $\phi^2 \bar{d}_i d_i$, and $\phi^2
\bar{d}_i i\gamma_5 d_i$ are considered, where the index $i$ runs over
the families. On the other hand, for the Majorana fermion DM $\chi$,
the nine operators $(\bar{\chi} \gamma^\mu \gamma_5 \chi) (\bar{u}_i
\gamma_\mu P_R u_i)$, $(\bar{\chi} \gamma^\mu \gamma_5 \chi)
(\bar{d}_i \gamma_\mu P_R d_i)$, and $(\bar{\chi} \gamma^\mu \gamma_5
\chi) (\bar{u}_i \gamma_\mu P_L u_i + \bar{d}_i \gamma_\mu P_L d_i)$
are of interest. As new physics would generally result in several
operators in the effective low-energy theory, putting limits on
individual operators in isolation should be regarded as only an
indicative exercise. Keeping this fact in mind, we simply assume that
the effective lagrangian has only one of the operators mentioned above
and derive the corresponding bounds.

\subsection{Cosmological limits}

The thermal relic abundance of the DM~\cite{Komatsu:2010fb} imposes an
important constraint.  A small annihilation cross section of the DM
implies a larger resultant abundance, leading possibly to the
over-closure of the universe. This immediately leads to a lower
limit on the coefficient of the operator from the cross section. Indeed, 
a consonance with the WMAP results would lead to an even more restrictive
bound.

For the scalar DM $\phi$ having interactions
$(v\phi^2/\sqrt{2}\Lambda^2) (c^{(R)} \bar{q} q + c^{(I)} \bar{q}
i\gamma_5 q)$, where $q$ represents a quark ($q = u_i, d_i$) with
$m_q$ being its mass, the annihilation cross section of the DM $\phi$
(times relative velocity $v_{\rm rel}$) is evaluated to be 
\begin{eqnarray}
\left.\sigma v_{\rm rel}\right|_{\phi\phi}
\simeq
\frac{3v^2}{2\pi\Lambda^4}
\left[
\left(c^{(R)}\right)^2 (1 - m_q^2/m_\phi^2)^{3/2}
+
\left(c^{(I)}\right)^2 (1 - m_q^2/m_\phi^2)^{1/2}
\right] + {\cal O}(\epsilon),
\nonumber
\end{eqnarray}
where $\epsilon = (s - 4m^2_\phi)/(4m^2_\phi)$. Using a
semi-analytical formula for the DM abundance~\cite{Gondolo:1990dk}, we
obtain a lower limit on the coefficient $\sqrt{c^{(R)}}/\Lambda$ or
$\sqrt{c^{(I)}}/\Lambda$.

On the other hand, for the Majorana fermion DM $\chi$ governed by the 
interaction $(2c/\Lambda^2) (\bar{\chi} \gamma_\mu \gamma_5 \chi)
(\bar{q} \gamma^\mu P_{R(L)} q)$, the annihilation cross section reads
\begin{eqnarray}
\left.\sigma v_{\rm rel}\right|_{\chi\chi}
\simeq
\frac{4c^2}{\pi \Lambda^4}
\left(4 m_\chi^2 \epsilon + 3 m_q^2/2 \right)
\sqrt{1 - m_q^2/m_\chi^2},
\end{eqnarray}
where $\epsilon = (s - 4m^2_\chi)/(4m^2_\chi)$ again. As in the case
of the scalar DM, we obtain a lower limit on the coefficient
$\sqrt{c}/\Lambda$ using the same semi-analytical formula.

\subsection{Astrophysical limit 1 (Direct detection)}

Direct detection experiments put a stringent limit on a Wilson
coefficient when the corresponding operator contributes to the
spin-independent cross section for the DM scattering off a nucleon.
In this paper, we adopt the results of the XENON100
collaboration~\cite{:2012nq}.  For the scalar DM $\phi$, as shown in
eqn.(\ref{eq: scalar int}), there are two types of interactions,
namely scalar and pseudo-scalar. Since the pseudo-scalar interaction
is not constrained, the operators of the scalar DM are limited only
through the scalar interaction.

Using the formula discussed in Ref.~\cite{SI scattering}, the
spin-independent scattering cross section between the DM and a nucleon
is given by the following formula when the DM couples to light quarks
($u$, $d$, and $s$ quarks):
\begin{eqnarray}
\sigma_{\rm SI} =
\frac{\left(c^{(R)}\right)^2 v^2 m_N^4 f_{Tq}^2}
{2 \pi \Lambda^4 (m_\phi + m_N)^2 m_q^2},
\end{eqnarray}
where $m_N \simeq$ 940 MeV is the nucleon mass, while $f_{Tq}$ is the
parameter determined by the hadron matrix element $\langle N | \bar{q}
q | N \rangle$. The values of this parameter for $u$, $d$, and $s$
quarks are $f_{Tu} \simeq$ 0.028, $f_{Td} \simeq$ 0.028, and $f_{Ts}
\simeq$ 0.009, respectively~\cite{Oksuzian:2012rzb}%
\footnote{These values, which are obtained by the lattice QCD calculations,
are consistent with the ones extracted from the $\pi$-$N$ scattering data
with the aid of the chiral perturbation theory~\cite{Alarcon:2011zs}.}.
On the other
hand, when the DM couples to heavy quarks ($c$, $b$, and $t$ quarks),
the formula of the spin-independent scattering cross section 
changes to 
\begin{eqnarray}
\sigma_{\rm SI} =
\frac{\left(c^{(R)}\right)^2 2 v^2 m_N^4 f_{TG}^2}
{729 \pi \Lambda^4 (m_\phi + m_N)^2 m_q^2}.
\label{eq:SI2}
\end{eqnarray}
The parameter $f_{TG}$ is determined by the hadron matrix element
$\langle N | G_{\mu \nu}^a G^{a \mu \nu} | N \rangle$ and its value is
evaluated to be $f_{TG} \simeq$ 0.9431 through the trace anomaly
relation, namely $f_{Tu} + f_{Td} + f_{Ts} + f_{TG} = 1$. The
essential reason why the scattering cross section depends on the gluon
hadron matrix element is that the DM can scatter off a nucleon
through an effective interaction with gluons inside the nucleon, a process
mediated by a one-loop diagram involving the heavy quark%
\footnote{For light quarks, contributions to the
scattering cross section from the loop-diagrams 
have a form different from that in eqn.(\ref{eq:SI2}) 
and are suppressed by the mass of light
quarks.}.

For a Majorana fermion, the charge radius vanishes identically, and
all interactions with quarks involve the axial-vector current, namely
$\bar{\chi} \gamma_\mu \gamma_5 \chi$. Only the spatial components of
the interaction survive at the non-relativistic limit, and the
corresponding quark interaction becomes\footnote{Note that $\langle N|
  \bar{q}\gamma^i q | N \rangle=0$ in the non-relativistic limit.}
$\bar{q} \gamma^i \gamma_5 q$.  These operators, therefore, contribute
only to the spin-dependent scattering cross section, but not to the
spin-independent one. Since the limit on the former cross section from
direct detection experiments turns out to be much weaker than that
obtained from the LHC experiments, we do not consider this constraint.

\subsection{Astrophysical limit 2 (Indirect detection)}

Several products from DM annihilations in the galactic halo, for
example, gamma-rays, positrons, electrons, anti-protons, and
anti-deuterons, are utilized to detect the DM. Among these, one of the
most reliable limits on the annihilation cross section of DM comes
from the indirect detection experiment through gamma-ray
observations. In particular, the observations from the galactic center
and Milky Way satellites currently give the most stringent limit. In
this article, we put a limit on each operator according to
Ref.~\cite{Hooper:2012sr}, in which a bound on the DM
annihilation cross section obtained by the gamma-ray observation from
the galactic center at the Fermi-LAT experiment is presented.

The differential gamma-ray flux in a direction $\psi$ originated in DM
annihilations at the region around the galactic center is given by
\begin{eqnarray}
\Phi(\psi) =
\frac{1}{4\pi} \frac{\sigma v_{\rm rel}}{2 m_{\rm DM}^2}
\left[ \frac{dN_\gamma}{dE} \right]
\int_{l.o.s} dl \rho^2(l),
\end{eqnarray}
where $\sigma$ is the annihilation cross section with $v_{\rm rel}$
being the relative velocity between the two incident DM particles,
each of mass $m_{\rm DM}$. The mass density of DM (the DM profile) at
the position $l$ is denoted by $\rho(l)$, and the integration is
performed along the observer's line-of-sight. Of the many profiles
discussed in the literature, we use the ``Cored ($R_c = 1$kpc)'' DM
profile as it gives the most conservative limit on the annihilation
cross section. The fragmentation function $dN_\gamma/dE$, which is
nothing but the gamma-ray spectrum produced per annihilation, has been
estimated using the PYTHIA code~\cite{Sjostrand:2006za} for each possible 
final state for the primary process, viz. $u\bar{u}$, $s\bar{s}$, $b\bar{b}$, $t\bar{t}$, etc.

Using the result of the Fermi-LAT experiment, model
independent upper limits on the bin-integrated 
quantity $(\sigma v_{\rm rel}/m_{\rm
  DM}^2) \times \int^{E_f}_{E_i} dE [dN_\gamma/dE]$ for various energy
bins, $(E_i, E_f) =$ (0.3, 1), (1, 3), (3, 10), and (10, 100) in GeV
unit, are presented in Ref.~\cite{Hooper:2012sr}. The upper limit on the
annihilation cross section, $\sigma v_{\rm rel}$, is therefore
obtained when the DM mass and the fragmentation function are
given. As the DM velocity at the present universe is three orders
of magnitude smaller than the speed of light, the product ($\sigma \, v_{\rm rel}$)  is dominated by the $s$-wave component. For the scalar DM,
this is given by
\begin{eqnarray}
\sigma v_{\rm rel} \simeq
\frac{3v^2}{2\pi\Lambda^4}
\left[
\left(c^{(R)}\right)^2 (1 - m_q^2/m_\phi^2)^{3/2}
+
\left(c^{(I)}\right)^2 (1 - m_q^2/m_\phi^2)^{1/2}
\right],
\nonumber
\end{eqnarray}
where the interaction of the DM is assumed to be
$(v\phi^2/\sqrt{2}\Lambda^2) (c^{(R)} \bar{q} q + c^{(I)} \bar{q}
i\gamma_5 q)$. On the other hand, for the Majorana fermion DM $\chi$,
the s-wave component of the annihilation cross section is given by 
\begin{eqnarray}
\sigma v_{\rm rel} \simeq
\frac{6 c^2 m_q^2}{\pi \Lambda^4} \sqrt{1 - m_q^2/m_\chi^2}.
\end{eqnarray}
where  the interaction Lagrangian is $(2c/\Lambda^2) (\bar{\chi}
\gamma_\mu \gamma_5 \chi) (\bar{q} \gamma^\mu P_{R(L)} q)$. Since the
annihilation cross section is proportional to the fermion mass squared
(on account of the helicity suppression), only the interactions involving
the second and third generation quarks are constrained by the indirect
detection experiment. \\

In this section, we have discussed the method to put limits on the DM
operators using cosmological and astrophysical observations. Obtained
limits on each operator are summarized in Fig.\ref{fig: Scalar DM}
for the real scalar DM and Fig.\ref{fig: Fermion DM} for the
Majorana fermion DM in section \ref{sec: results}, where limits from
cosmology, direct detection, and indirect detection experiments are
shown as brown, green, and blue lines in both figures. Physical
interpretation of the results are discussed in section \ref{sec:
  results} in comparison with the limits obtained from the LHC
experiment.

\section{Constraints from the LHC experiment}
\label{sec: LHC}

We first discuss possible signals of DM at the LHC experiment. DM
particles are pair-produced from quark-quark interactions through
the higher dimensional operators discussed in section \ref{sec:
  operators}. However, exclusive pair production does not result in a
visible object to tag on to  and, hence, cannot be detected by the
experiment. We, thus, need to pair-produce DM particles in
association with jets or photons. In this case, the DM pair recoils
against the visible particle(s) and produces 
missing transverse energy ($\slashed{E}_T$). The
generic signature of DM production at the LHC experiment is thus
$\slashed{E}_T$ plus a few number of jets or photons.

In the context of LHC searches, we can divide the DM effective
operators into three categories, i.e., the operators containing light
quarks, bottom quarks and top quarks respectively.
In the first
case, we can pair-produce DM particles with gluon/quark radiations
from initial quark/gluon legs and thus observe the signal in the
mono-jet plus $\slashed{E}_T$ channel. In the second case, we need two
bottom quarks in the initial state that may come from the gluon
splitting\footnote{Note that a gauge-invariant calculation would also
  need to included diagrams such as $g b \to b \phi \phi$ with the
  initial $b$ emanating from gluon spitting.}. This means, we may
produce DM pair alongwith two b-jets. The DM interaction with top
quarks is very different from the other cases. Here we need to produce
two DM particles in association with two top quarks and the signal
looks like top pair plus $\slashed{E}_T$. This final state is
particularly interesting as it comes from different beyond the SM
processes.

From the  discussions above, it is clear that the signals will comprise
events with $\slashed{E}_T$ plus a few of SM particles with the hope that 
the shape of the $\slashed{E}_T$ distribution may be used to distinguish 
these from the SM backgrounds. However, unlike the R-parity conserving
SUSY, $\slashed{E}_T$ comes from the recoil of jets against DM
particles and thus produces featureless rapidly decreasing
distribution for DM coupled to light and bottom quarks. In case of top
quark--DM effective operators, the situation is slightly different as
$\slashed{E}_T$ can also come from leptonic decays of $W$
bosons. It is therefore important to compare the $\slashed{E}_T$
distributions for these two cases. The $\slashed{E}_T$ distribution
becomes harder with increasing the DM mass although the overall
production cross section decreases with the mass.

\begin{figure}[t]
\begin{center}
\includegraphics[width=0.8\hsize]{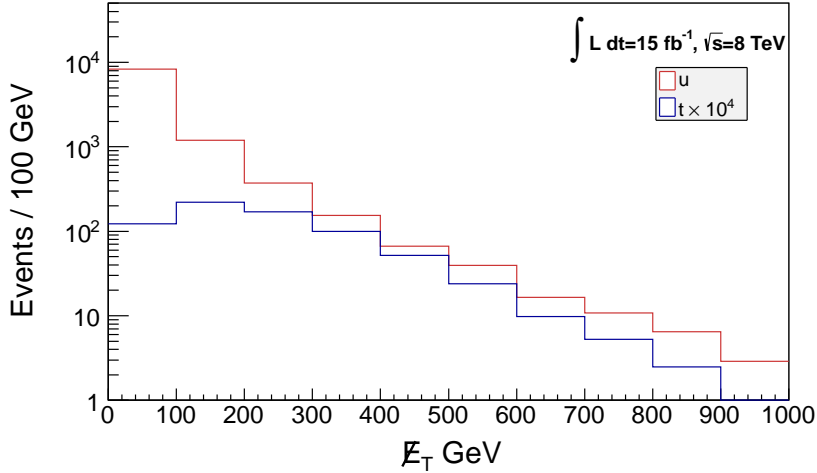}
\caption{\sl \small $\slashed{E}_T$ distributions for the two operators described in the text at 8 TeV run.}
\label{fig: monojUL-7tev}
\end{center}
\end{figure}

For illustrative purposes, consider the two operators $(2\bar{\chi}
\gamma^\mu \gamma_5 \chi) (\bar{u} \gamma_\mu P_R u)/\Lambda^2$ and
$(2\bar{\chi} \gamma^\mu \gamma_5 \chi) (\bar{t} \gamma_\mu P_R
t)/\Lambda^2$ assuming $m_{\chi}$ = 100 GeV and $\Lambda =$ 1 TeV. The
corresponding $\slashed{E}_T$ distributions are shown in Fig.\ref{fig:
monojUL-7tev} for the 8 TeV run with 15 fb$^{-1}$ integrated
luminosity. It is immediately apparent that the $\slashed{E}_T$
distribution for the top-quark case is slightly harder compared to the
light quark case. However, with the cross section being much smaller,
it is not possible to probe a DM that couples primarily to top quarks
(and with a strength comparable to the electroweak interactions) from
the current (8 TeV) run and we have to consider high luminosity LHC
run (100 fb$^{-1}$) with $\sqrt{s} =$ 14 TeV.

\subsection{Model independent upper limits}

We now discuss upper limits on the cross section
times acceptances for aforementioned interactions.
We consider three specific 
channels, namely mono jet + $\slashed{E}_T$, mono b-jet + $\slashed{E}_T$ and
top quark(s) + $\slashed{E}_T$ searches. In order to estimate the
prospects, we perform 
Monte-Carlo (MC) simulations. Madgraph5~\cite{Alwall:2011uj} is
utilized to generate parton level events and the events are interfaced
to PYTHIA6.420~\cite{Sjostrand:2006za} to deal with showering and
hadronization. For detector simulation, we utilize
Delphes2.0.2~\cite{Ovyn:2009tx},
which takes into account effects of mis-measurements in a MC way,
 and resolution parameters are chosen
based on the CMS detector performance given in
Ref.~\cite{Chatrchyan:2011nd}.

\subsubsection{Mono jet + $\slashed{E}_T$}
\label{sec: mono_upper}
The upper limit on the non-standard model contribution to the mono jet
+ $\slashed{E}_T$ channel have already been reported in
Refs.~\cite{Chatrchyan:2011nd, Chatrchyan:2012pa, ATLAS-monojet}
for the 7 TeV run. In this paper, we take the results of
Ref.~\cite{Chatrchyan:2012pa} and set the constraints. We also
estimate the prospect for 8 TeV run by the MC simulation imitating the
search discussed in the reference. The basic selections used in the
analysis are the following:
\begin{enumerate}
\item Event must have a missing transverse energy $\slashed{E}_T>$ 200 GeV.
\item The leading jet must have $p_T > $110 GeV and $|\eta| <$ 2.4.
\item The number of jets with $p_T >$ 30 GeV needs to be  smaller than 3.
\item There should be no isolated leptons or tracks with $p_T >$ 10 GeV.
\item The second jet must have $p_T <$ 30 GeV or the difference of the azimuthal \\
 angle between the leading and the second jets should be smaller than 2.5.
\end{enumerate}
Here, $p_T$ and $\eta$ denote the transverse momentum and
pseudo-rapidity respectively. In Ref.~\cite{Chatrchyan:2012pa}, the
upper limit on the number of the non-standard model events has been
obtained, when an additional cut of $\slashed{E}_T>$ 250, 300, 350, or
400 GeV is applied. We express the results in terms of the cross
section times the acceptance ($\sigma \times {\cal A}$), which are
summarized in the table shown below;

\begin{center}
\begin{tabular}{c|cccc}
Cut on $\slashed{E}_T$ (GeV) & 250 & 300 & 350 & 400 \\
\hline
$\sigma \times {\cal A}$ (fb) & 120 & 73.6 & 31.6 & 19.0 \\
\hline
\end{tabular}
\end{center}

In order to estimate the prospect with the 8 TeV run as accurately as
possible, we first perform the MC simulation for a 7 TeV run and
compare the background numbers with those of
Ref.~\cite{Chatrchyan:2012pa} and obtain normalization factors
required to adjust the difference. The same normalization factors are
used to calculate the backgrounds for the 8 TeV run\footnote{This
approximation is expected to be quite an accurate one as the modest
increase in the operating energy does not change the features of the
distributions to an appreciable degree.}. Note that the cross section
obtained by MadGraph and PYTHIA is only at LO, and one needs a K factor
to obtain the correct normalization. Instead using the K factor, we
compare our simulation to the distribution in the reference. As is
shown in the reference, $W \to l\nu$ + jets, $Z \to \nu\nu$ + jets and
$t\bar{t}$ + jets are the dominant backgrounds, where $l$ denotes a
charged lepton. Other possible sources such as QCD, single top,
etc. have negligible contributions. The cut flow for the standard
model events is shown in Table~\ref{table: monoj-cutflow} assuming an
integrated luminosity of 15 fb$^{-1}$.

\begin{table}[t]
\begin{center}
\begin{tabular}{c|ccc}
& $W\rightarrow l\nu$+jets & $Z\rightarrow \nu\nu$+jets& $t\bar{t}$+jets \\
\hline
Cut 1-3                 & 176369 & 62634 &  4756 \\
Cut 4                   & 157616 & 58216 &  4016 \\
Cut 5                   &  39084 & 56597 &   625 \\
$\slashed{E}_T>250$ GeV &  12915 & 21986 &   193 \\
$\slashed{E}_T>300$ GeV &   4950 &  9588 &    59 \\
$\slashed{E}_T>350$ GeV &   2065 &  4674 &    28 \\
$\slashed{E}_T>400$ GeV &    928 &  2337 &    20 \\
$\slashed{E}_T>450$ GeV &    473 &  1275 &     8 \\
$\slashed{E}_T>500$ GeV &    223 &   754 &     7 \\
\hline
\end{tabular}
\caption{\sl \small Cut flow table for the SM backgrounds at the mono jet + $\slashed{E}_T$ channel. The center of mass energy of 8 TeV and the integrated luminosity of 15 fb$^{-1}$ are assumed.}
\label{table: monoj-cutflow}
\end{center}
\end{table}

We have calculated the expected 95\% C.L. upper limit on the
non-standard model cross section times the acceptance, which is
summarized in the table below. Here, we have assumed that the
systematic uncertainty associated with the estimation of the SM
background events is 10\%, which is same as the one in
Ref.~\cite{Chatrchyan:2012pa}.

\begin{center}
\begin{tabular}{c|cccccc}
Cut on $\slashed{E}_T$ (GeV) & 250 & 300 & 350 & 400 & 450 & 500 \\
\hline
$\sigma \times {\cal A}$ (fb) & 459& 191 & 89.1 & 43.6 & 23.7 & 13.7 \\
\hline
\end{tabular}
\end{center}

\subsubsection{Mono b-jet + $\slashed{E}_T$}

The mono b-jet + $\slashed{E}_T$ channel is expected to be useful to
search for a DM that couples   mainly to bottom quarks. In addition
to the selection cuts used in the previous subsection (the cuts 1--5),
we also require 
\begin{enumerate}
\setcounter{enumi}{5}
\item The leading jet is b-tagged.
\end{enumerate}
For the b-tagging method, we assume the following properties:
\begin{itemize}
\item The b-tagging efficiency is 0.6 for real b jets.
\item The mis-tagging rate for light jet is 0.004.
\item The mis-tagging rate for c-jet is 0.1.
\end{itemize}
These tagging efficiencies can be achieved in the CSVM tagger
calibrated in Ref.~\cite{CMS-btag}. With the b-jet being the leading
one, a large $\slashed{E}_T$ would, typically, require a large $p_T$
for the b-jets. This, though, would lower the efficiency of
b-tagging.  from the value quoted above. Hence, we do not raise the
$\slashed{E}_T$-cut to levels as high as those for the case of the
mono jet + $\slashed{E}_T$ channel, and just impose a moderate
requirement of $\slashed{E}_T > 350$ GeV.

The estimation of the number of the SM events after the cuts 1--6 and
$\slashed{E}_T >$ 350 GeV is shown in Table~\ref{table:
  monobj-cutflow} for 8 TeV run assuming 15 fb$^{-1}$ data. Here,
``jets'' also include those originating in heavy flavors. Since the
mono b-jet + $\slashed{E}_T$ channel is almost similar to the mono jet
+ $\slashed{E}_T$ channel, the dominant backgrounds in the previous
subsection are also the dominant ones here. We use the same
normalization factor as the one in the mono jet + $\slashed{E}_T$
channel to estimate the number of the events.

\begin{table}[t]
\begin{center}
\begin{tabular}{c|ccc}
& $W\rightarrow l\nu$+jets & $Z\rightarrow \nu\nu$+jets& $t\bar{t}$+jets \\
\hline
$\slashed{E}_T>350$ GeV & 33 & 111 & 23 \\
\hline
\end{tabular}
\caption{\small \sl Expected number of the SM events in the mono b-jet
  + $\slashed{E}_T$ channel. A center of mass energy of 8 TeV and an
  integrated luminosity of 15 fb$^{-1}$ are assumed.}
\label{table: monobj-cutflow}
\end{center}
\end{table}

The expected 95\% C.L. upper limit on the cross section times the
acceptance for the non-SM event is estimated to be {\bf 2.9 fb}. Here,
we have again assumed that the systematic uncertainty in the
estimation of the SM events is 10\%. One may expect that the use of
tighter b-tagging condition would efficiently reduce the $W$ and $Z$
backgrounds. However, it turns out that half of the remaining $W$ and
$Z$ events include the real b-jets with the b-tagging properties we
adopt. Therefore, tightening the b-tagging condition does not improve
the result of our analysis.

\subsubsection{Top quark(s) + $\slashed{E}_T$}

The case of the DM coupling primarily to the top quark needs separate
treatment. At the LHC, such a DM can only be pair-produced in
association with two top quarks, and the production cross section is
highly suppressed in comparison to the case of light quarks. To
compensate for this, we concentrate on the prospects for the 14 TeV
run, rather than the the 7(8) TeV ones.

The top quark decays promptly to a hard b-jet and an on-shell
$W$. If the $W$, in turn, decays into a lepton-neutrino pair, the
last-mentioned would contribute to missing momentum. In other words,
both the semileptonic and dileptonic decay channels for the $t \bar t$
pair are associated with an inherent $\slashed{E}_T$, and recognizing
the additional missing momentum due to the DM pair would be a
nontrivial process. Indeed for such channels (i.e., semileptonic or
dileptonic decays), the SM $t \bar t$ production itself proves to be
an almost insurmountable source of background to the DM signal.  In
order to suppress this, we consider instead the pure hadronic mode
(even with all the associated complications).  We, therefore, impose
the following basic selection cuts:
\begin{enumerate}
\item There is no isolated light lepton ($e^\pm, \mu^\pm$) with $p_T >$ 20 GeV.
\item There are at least four jets with $p_T >$ 100, 80, 50, 50 GeV
(sequentially) \\ and one of them has to be b-tagged.
\item The azimuthal angles between the missing transverse \\ momentum
and the leading four jets are larger than 0.2.
\item $M_{\rm eff} >$ 1000 GeV and $\slashed{E}_{T}/M_{{\rm eff}} >$ 0.3.
\end{enumerate}
Here, $M_{\rm eff}$ is the scalar sum of the transverse momenta of the
leading 4 jets and $\slashed{E}_T$.

\begin{figure}[!h]
\begin{center}
\includegraphics[width=0.8\hsize]{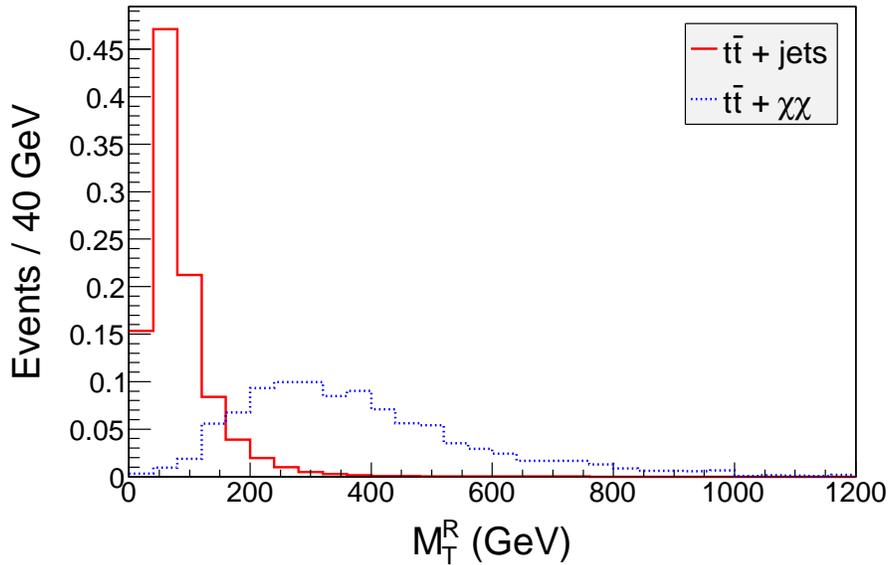}
\caption{\sl \small Normalized distribution of the $M_T^R$ variable
for the SM background ($t\bar{t}$) and for DM effective operator
$(2\bar{\chi} \gamma^\mu \gamma_5 \chi) (\bar{t} \gamma_\mu P_R
t)/\Lambda^2$ events after imposing the selection cut 1--3. Center of
mass energy of the LHC experiment is set to be 14 TeV.}
\label{fig: mrt}
\end{center}
\end{figure}
The top quark(s) + $\slashed{E}_T$ channel resembles the supersymmetry
search with multi-jet + $\slashed{E}_T$~\cite{Aad:2009wy}. The
dominant backgrounds accrue from $W+ {\rm jets}$ (with $W\to l\nu$),
$Z + {\rm jets}$ (with $Z\to \nu\nu$) and $t\bar{t} + {\rm
jets}$.
Interestingly, we find that the $M_T^R$
variable~\cite{Rogan:2010kb} is useful in separating the signal events
from the SM $t\bar{t}+{\rm jets}$ events. In order to define the
variable, we first define two mega jets. If $n$ jets are present in
the event, $2^{n-1}$ different combinations 
are possible in such a reconstruction.
Of these, we choose the one which minimizes the sum of the
squares of the two mega jet masses. From the four momentum of the two
mega jets $p^{j1}$, $p^{j2}$ and the transverse missing momentum
$\vec{\slashed{E}}_T$, the quantity $M_T^R$ is defined by
\begin{eqnarray}
M_T^R =
\sqrt{\frac{\slashed{E}_T(p_T^{j1}+p_T^{j2})
-\vec{\slashed{E}}_T \cdot (\vec{p}_T^{j1}+\vec{p}_T^{j2})}{2}} \ .
\end{eqnarray}
For the SM $t\bar{t}$ events, the distribution
falls sharply for $M_T^R \gsim m_t$, while that for the signal is expected to be
broader. We show the $M_T^R$ distributions for the SM $t\bar{t}$
and for the DM effective operator $(2\bar{\chi} \gamma^\mu \gamma_5 \chi)
(\bar{t} \gamma_\mu P_R t)/\Lambda^2$ in Fig.~\ref{fig: mrt}
assuming $m_{\chi}$ = 100 GeV and $\Lambda =$ 1 TeV after imposing the
selection cuts 1--3. Based on this distribution, we impose additional
cuts $M_T^R >$ 300, 400, 500, 600, 700, 800 and 900 GeV. The cut flow
for the SM backgrounds, namely $t\bar{t}$ +jets, $W$ + jets,
and $Z$ + jets backgrounds are shown in Table~\ref{table:
  top-cutflow}.

\begin{table}[t]
\begin{center}
\begin{tabular}{c|c|c|c}
& $t\bar{t}$+jets & $Z (\rightarrow \nu\nu)$+jets& $W (\rightarrow l\nu)$+jets\\
\hline
Cut 1-4          & 5746     & 258 & 176 \\
$M_T^R>$ 300 GeV & 5723     & 255 & 176 \\
$M_T^R>$ 400 GeV & 4753     & 239 & 174 \\
$M_T^R>$ 500 GeV & 1934     & 200 & 141 \\
$M_T^R>$ 600 GeV & 659      & 143 & 95 \\
$M_T^R>$ 700 GeV & 239      & 85  & 50 \\
$M_T^R>$ 800 GeV & 94       & 47  & 19 \\
$M_T^R>$ 900 GeV & 21       & 21  & 14 \\
\hline
\end{tabular}
\caption{\sl \small Expected cut flow for the SM backgrounds in the top + $\slashed{E}_T$ channel. The center of mass energy of 14 TeV and the integrated luminosity of 100 fb$^{-1}$ are assumed.}
\label{table: top-cutflow}
\end{center}	
\end{table}

As expected, the $t\bar{t}$+ jets background reduces drastically with stronger
$M_T^R$ cuts, whereas the $Z$ and $W$ backgrounds are not reduced
significantly. However, given that the latter are inherently smaller, this 
is not worrisome. Here, the number of the $t\bar{t}$ events
is normalized assuming the total cross section of 920
pb~\cite{Moch:2008qy}. The numbers of the $W \to l\nu$ + jets and $Z
\to \nu\nu$ + jets events are normalized with the matched cross
section obtained from the MadGraph-PYTHIA package. We have calculated
the expected 95\% C.L. upper limit on the non-standard model cross
section times the acceptance, which is shown in the table below. Here,
the systematic uncertainty in the estimation of the standard model
events was assumed to be 10\%, which is typical of supersymmetry
searches with the hadronic mode.

\begin{center}
\begin{tabular}{c|ccccccc}
Cut on $ M_T^R$ (GeV) & 300 & 400 & 500 & 600 & 700 & 800 & 900 \\
\hline
$\sigma \times {\cal A}$ (fb) & 12.2 & 10.2 & 4.57 & 1.87 & 0.839 & 0.418 & 0.208 \\
\hline
\end{tabular}
\end{center}

In this paper, only the result for the hadronic mode has been
shown. On the other hand, we also have estimated the prospects using
the one (semileptonic) and two lepton (leptonic) modes. For the
leptonic mode, we use the $M_{T2}$ variable constructed from
$\slashed{E}_T$ and the transverse momentum of two leptons. For the
semi-leptonic mode, the $M_T$ variable constructed from
$\slashed{E}_T$ and the transverse momentum of a lepton are used to
reduce the background. These methods are discussed in
Ref.~\cite{Plehn:2012pr} in  the
context of scalar top searches. We find, however, that the corresponding 
constraints  are weaker than those obtained with the hadronic mode. \\

We are now ready to discuss the imposition of 
constraints on each operator from the LHC experiment. For this
purpose, we implement the interactions described in section \ref{sec:
  operators} in Madgraph5 with the aid of
Feynrules~\cite{Christensen:2008py} and perform MC simulations for
three channels as discussed above.

\begin{figure}[t]
\begin{center}
\includegraphics[scale=0.53]{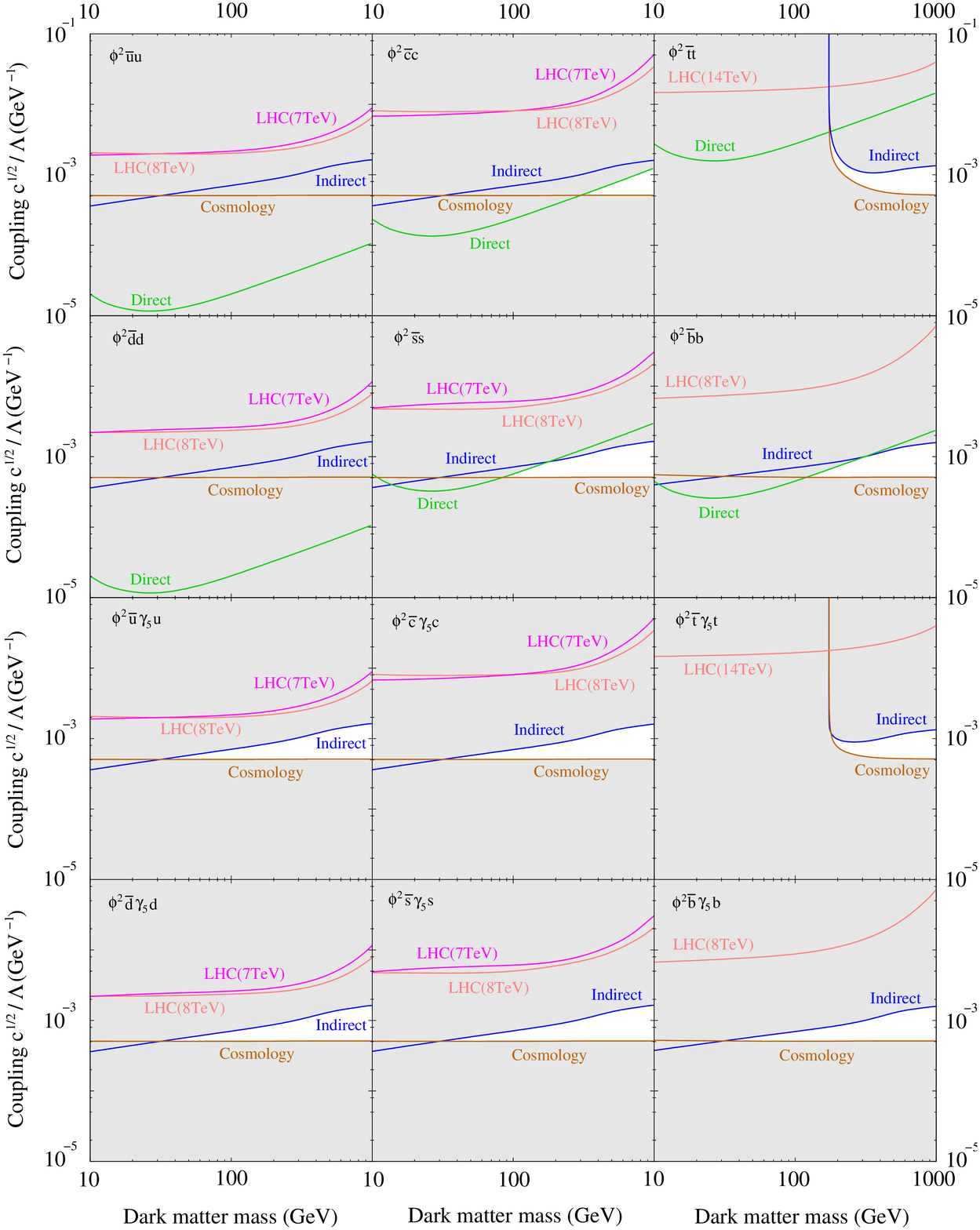}
\caption{\small \sl Limits on the operators describing the
  interactions between the real scalar DM $\phi$ and various
  quarks. Those from current LHC `LHC(7TeV)', future LHC `LHC(8TeV) \&
  LHC (14TeV)', cosmological `Cosmology', direct detection `Direct',
  and indirect detection `Indirect' experiments are shown as magenta,
  pink, brown, green, and blue lines, respectively in each panel of
  the figure.}
\label{fig: Scalar DM}
\end{center}
\end{figure}

In order to put constraints on the operators involving light quarks
(up, down,~charm and strange quarks), we perform the mono jet
$+\slashed{E}_T$ analysis. We find the optimized cut on
$\slashed{E}_T$ by maximizing the ratio of the efficiency for
accepting the signal to the upper bound on the cross section times the
acceptance for the non-standard model events. As a reference point, 
we choose the DM mass of 100 GeV and
the interaction with up quark.\footnote{The optimized values do not
  change by more than 50 GeV even if we choose other reference points.}
The results are summarized in Table~\ref{table: optimization
  monoj}. As expected, a stronger cut on $\slashed{E}_T$ gives more
severe bound on the interactions for the Majorana fermion DM. It is
possible that a cut on $\slashed{E}_T$ even stronger than those discussed 
in section \ref{sec: mono_upper} would result in a more severe
bound. However, such strong cuts typically lead to larger systematic errors, 
and, hence, we do not consider this avenue. The
current and future expected constraints on each operator are shown in
the first and the second columns of Fig.\ref{fig: Scalar DM} and
\ref{fig: Fermion DM} in the next section.

\begin{table}[t]
\begin{center}
\begin{tabular}{c|cc}
& Real scalar & Majorana fermion \\
\hline
7 TeV & 350 GeV & 400 GeV \\
8 TeV & 500 GeV & 500 GeV \\
\hline
\end{tabular}
\caption{\sl \small Optimized cut on $\slashed{E}_T$ for the real scalar and the Majorana fermion DMs.}
\label{table: optimization monoj}
\end{center}
\end{table}

In order to put constraints on the operators involving bottom and top
quarks, we perform the mono b-jet $+ \slashed{E}_T$ and the top
quarks(s) $+ \slashed{E}_T$ analysis, respectively. For the top
quarks(s) $+ \slashed{E}_T$ analysis, the optimized cut on $M_T^R$
turns out to be 900 GeV for both the real scalar and the Majorana
fermion DMs. The future expected constraints on the operators are
shown in the third columns of Figs.~\ref{fig: Scalar DM} and
\ref{fig: Fermion DM}.

\section{Results}
\label{sec: results}

We have already explained
in the preceding two sections that the limits come mainly from WMAP,
Fermi-LAT, XENON100 and LHC experiments. Since the structures of the
operators are very different for the real scalar and the Majorana
fermion DMs, we discuss the results of these two classes of
interactions separately. In all cases, we
vary the DM mass from 10 GeV to 1 TeV and calculate the limits and future
prospects in terms of the only relevant
combination of the cut-off scale ($\Lambda$)
and the coupling ($c$), namely $\sqrt{c}/\Lambda$.

\subsection{Scalar DM}

All the (expected) limits for the real scalar DM
interactions are summarized in Fig.\ref{fig: Scalar
  DM}, where constraints from cosmology, direct detection, and indirect
detection experiments are shown as brown, green, and blue lines, while
the current limit and future prospect from the LHC experiment are
shown as magenta and pink lines, respectively. It can be seen that the
limits obtained from the indirect detection are much stronger than
those from the LHC experiment except for the operators $\phi^2
\bar{t}t$ and $\phi^2\bar{t}\gamma^5t$. This is because the production
cross section at the LHC experiment is small for the scalar DM, while
its annihilation cross section is large as there is neither helicity
nor p-wave suppression.

The limit from the direct detection experiment is also very strong for
the scalar interactions, since they contribute to the spin-independent
scattering cross section. The limit on the operator $\phi^2\bar{c}c$
is stronger than that on $\phi^2\bar{s}s$. This is so 
as the hadronic matrix element $\langle N| \bar{s}s |N \rangle$
is small ($f_{T_s} \simeq 0.009$), while the interaction with $c$
quark is expressed by the interaction with gluons at low energy and
the corresponding matrix element $\langle N | G_{\mu \nu}^a G^{a \mu \nu} | N
\rangle$ is large ($f_{T_G} \simeq 0.9431$).

It is worth noting that the lines obtained from the cosmological limit
are almost independent of the DM mass
 except for the top quark mass threshold for the
operators $\phi^2 \bar{t}t$ and $\phi^2\bar{t}\gamma^5t$. This is
so because the annihilation cross section of the scalar DM is controlled
by the dimension-5 operator after the electroweak symmetry breaking,
and the cross section depends very weakly on the DM mass.

It is also clear that the scalar interactions with $u$ and $d$ quarks
are not favored from current data, whereas the interaction with top
quark is the most difficult one to constrain. Even with 100 fb$^{-1}$
data at 14 TeV run, it would be challenging to cover most of the
parameter space consistent with WMAP data.

\subsection{Fermion DM}

\begin{figure}[t]
\begin{center}
\includegraphics[scale=0.53]{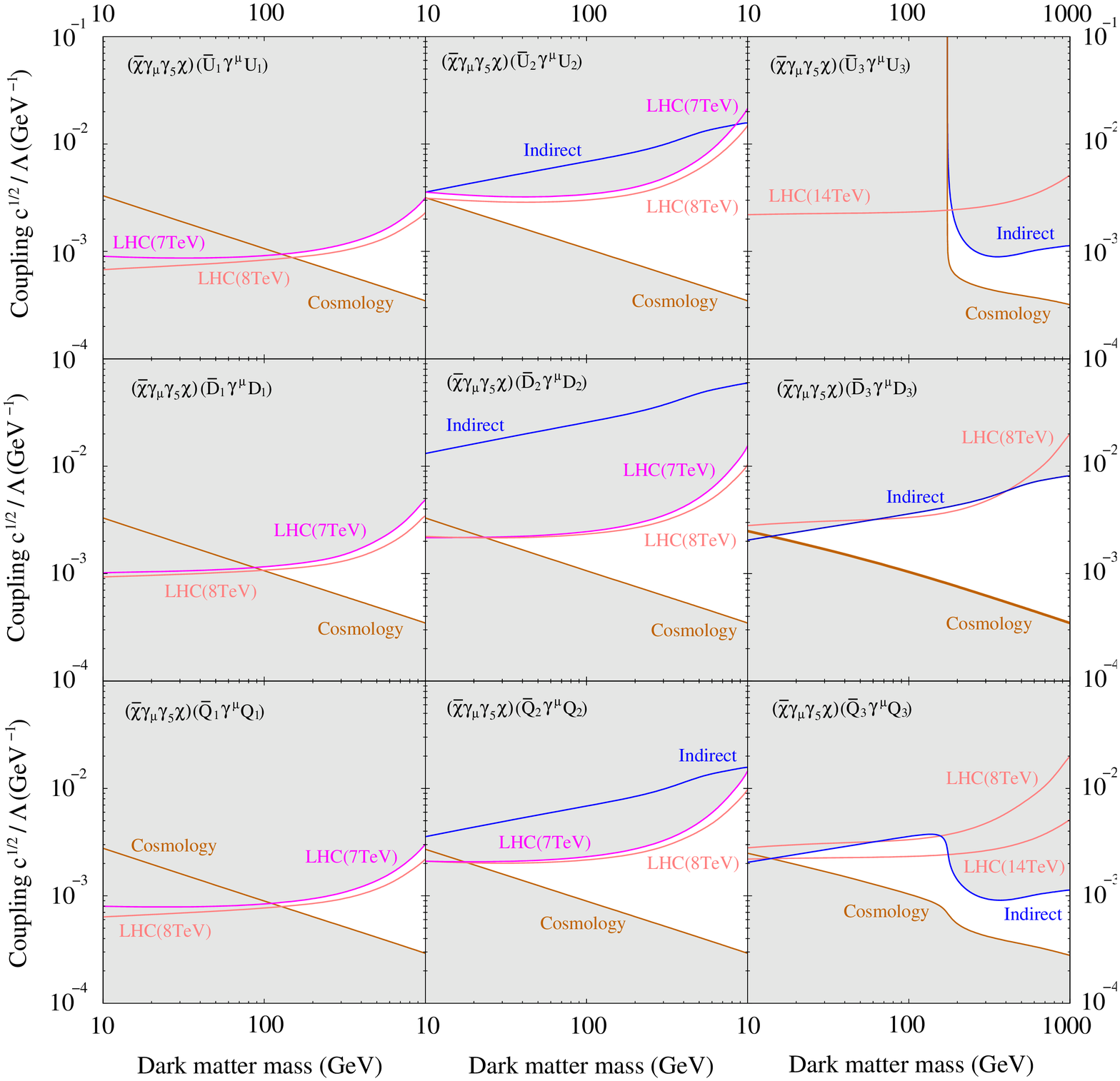}
\caption{\small \sl Limits on the operators describing the interaction
  between the Majorana fermion DM $\chi$ and various quarks. Those
  from current LHC `LHC(7TeV)', future LHC `LHC(8TeV) \& LHC (14TeV)',
  cosmological `Cosmology', direct detection `Direct', and indirect
  detection `Indirect' experiments are shown as magenta, pink, brown,
  green, and blue lines, respectively in each panel of the figure.}
\label{fig: Fermion DM}
\end{center}
\end{figure}

For the Majorana fermion DM, all 
results are summarized in Fig.\ref{fig: Fermion DM}, where the
(expected) limits are shown in the same way as for the scalar DM case. 
A remarkable
point is that the limit from the indirect detection experiment is weak
due to the helicity suppression. It is to be noted that the limit from
the direct detection experiment is also very weak (and not shown in
the figure), because the operators do not contribute to the
spin-independent scattering cross section. It is therefore important
to investigate the experimental way to search for the Majorana fermion
DM at the LHC experiment.

The mono jet $+\slashed{E}_T$ search gives strong limits on the DM
interactions with light quarks (up, down, charm and strange quarks) in
comparison with the limits obtained from the indirect detection
experiment in the most of the parameter space, and so will the mono
b-jet + $\slashed{E}_T$ search on the interaction with bottom
quark. Unfortunately, the limit from the 8 TeV run will not be much
stronger than the one obtained in the 7 TeV run despite the increase in signal 
cross section, as the number of
SM background events increases as well.
On the other hand, the top quark(s) + $\slashed{E}_T$ search
will give comparable limits to those from the indirect detection for
the interaction with top quark, which is in sharp contrast to the case
of the real scalar DM.

\section{Summary}
\label{sec: summary}

We systematically investigate DM interactions with various quarks in
the framework of effective operators. We consider the DM to be a
singlet scalar or a Majorana fermion and thoroughly study bounds and
future prospects of the DM discoveries from different astrophysical,
cosmological and collider experiments. We have shown that the mono
b-jet + $\slashed{E}_T$, and the top quark(s) + $\slashed{E}_T$
channels along with the mono jet + ($\slashed{E}_T$) channel could be
very important to search for the DM interacting with quarks. Here, we
derive model-independent upper limits on the cross section times
acceptance for these three channels mentioned above at 7 TeV (8 or 14
TeV) run of the LHC experiment. It is to be noted that the model
independent bounds derived in our paper can also be used to constrain
other DM or new physics models with similar type of signatures.

For the scalar DM with spin-independent interactions, the direct
detection experiment supersedes the LHC bound because of smallness of
the production cross section at the LHC experiment. Irrespective of 
whether it is spin-independent interaction or not, the indirect
detection limit from Fermi-LAT data on the scalar DM is found to be
stronger than the LHC bound. However, in case of the Majorana DM
interactions considered in our work, the LHC experiment plays very
important role for detection or exclusion of such possibility as the
spin-independent cross section is zero. Possibility of detecting the
top quark interaction with the DM is unfortunately not so promising at
the LHC experiment even with 100 fb$^{-1}$ data at 14 TeV run.

\section*{Acknowledgments}

This work is supported by the Grant-in-Aid for Scientific research
from the Ministry of Education, Science, Sports, and Culture (MEXT),
Japan (Nos. 22244021, 23740169 for S.M., \& Nos. 22540300, 23104006
for M.M.N.), and also by the World Premier International Research
Center Initiative (WPI Initiative), MEXT, Japan. DC thanks the
Department of Science and Technology, India for assistance under the
project DST-SR/S2/HEP-043/2009 and acknowledges partial support from
the European Union FP7 ITN INVISIBLES (Marie Curie Actions,
PITN-GA-2011-289442). The work of K.H. is supported by JSPS Research
Fellowships for Young Scientists.


\end{document}